\documentclass[aip,jcp,unsortedaddress,reprint,a4paper,amsmath,amssymb,]{revtex4-1}

\usepackage{graphicx,color}
\usepackage{epstopdf}
\usepackage{dcolumn}

\newcommand{\Op}[1]{\boldsymbol{\mathsf{\hat{#1}}}}

\def\openone{\leavevmode\hbox{\small1\kern-3.3pt\normalsize1}}

\begin{document}

\title{A Chebychev propagator with iterative time ordering for
  explicitly time-dependent Hamiltonians}
\date{\today}

\author{Mamadou Ndong}
\affiliation{Institut f\"ur Theoretische Physik,
  Freie Universit\"at Berlin,
  Arnimallee 14, 14195 Berlin, Germany}

\author{Hillel Tal-Ezer}
\affiliation{School of Computer Sciences, The Academic College of
  Tel Aviv-Yaffo,  Rabenu Yeruham St., Tel-Aviv 61803, Israel} 

\author{Ronnie Kosloff}
\affiliation{Institute of Chemistry and
  The Fritz Haber Research Center,
  The Hebrew University, Jerusalem 91904, Israel
  }

\author{Christiane P. Koch}
\email{ckoch@physik.fu-berlin.de}
\affiliation{Institut f\"ur Theoretische Physik,
  Freie Universit\"at Berlin,
  Arnimallee 14, 14195 Berlin, Germany}

\begin{abstract}
  A propagation method for time-dependent 
  Schr\"odinger equations with an explicitly time-dependent
  Hamiltonian is developed where time ordering is achieved iteratively.  
  The explicit time-dependence of the time-dependent Schr\"odinger
  equation is rewritten as an
  inhomogeneous term. At each step of the iteration, the resulting
  inhomogeneous Schr\"odinger equation is solved with the Chebychev
  propagation scheme presented in J. Chem. Phys. 130, 124108
  (2009). The iteratively time-ordering Chebychev propagator is shown to
  be robust, efficient and accurate and compares very favorably to all
  other available propagation schemes.
\end{abstract}

\maketitle


\section{Introduction}
\label{sec:intro}

The dynamics of the interaction of matter with a strong radiation
field is described by  
time-dependent Schr\"odinger equations (TDSEs) where the Hamiltonian is
explicitly time-dependent.  This description is at the core of the theory of
harmonic generation,\cite{Corkum94,Corkum05}  pump-probe
spectroscopy,\cite{engel98} 
and coherent control.\cite{k67,ZhuJCP98}  
Typically, an atom or molecule couples to a laser pulse via a
dipole transition, 
\begin{equation}
  \label{eq:H}
  \Op{H}(t) = \Op{H}_0 + E(t) \Op{\mu}\,,
\end{equation}
with $E(t)$ the time-dependent electromagnetic field, causing the explicit
time-dependence of the Hamiltonian.  Simulating these light-matter
processes from first principles 
imposes a numerical challenge. Realistic simulations require efficient
procedures with very high accuracy. 

For example, in coherent control processes, 
interaction of quantum matter with laser light leads to
constructive interference in some desired channel and destructive
interference in all other channels. In time-domain coherent control
such as pump-probe spectroscopy, wave packets created by radiation 
at an early time interfere with wave packets generated at a later time.
This means that the relative phase between different partial wave
packets has to be maintained for long time with high accuracy. As a
result, numerical methods designed to simulate such 
phenomena have to be highly accurate, minimizing the errors in both amplitude and phase.

The difficulty of simulating explicitly time-dependent Hamiltonians,
emerges from the fact that the commutator of the 
Hamiltonian with itself at different times does not vanish,\cite{magnus09}
\begin{equation}
  \label{eq:Ht1t2}
  [\Op{H}(t_1),\Op{H}(t_2)]_- \neq 0\,.
\end{equation}
Formally, this effect is taken into account by time ordering such that
the time evolution is given by
\begin{equation}
  \label{eq:timeordering}
  \Op{U}(T,0) = \mathcal{T} e^{-\frac{i}{\hbar}\int_0^T 
    \Op{H}(t)\,dt}\,. 
\end{equation}
The effect of time ordering is to incorporate higher order commutators
into the propagator $ \Op U (T,0) $. 
For strong fields $E(t)$ and fast time-dependences the
convergence with respect to ordering is slow.  
Methods to incorporate the second order Magnus term\cite{magnus} have
been developed 
either in a low order polynomial expansion\cite{k79,k92} or as a
split exponential.\cite{klaiber09}

A quantum dynamical propagator that fully accounts for time ordering
is given by the $(t,t')$ method.\cite{Ronniettprime} It is based on
rewriting the Hamiltonian in an extended Hilbert space where an
auxiliary coordinate, $t'$, is added and terms such as $E(t')\Op{\mu}$
are treated as a potential in this degree of freedom. The Hamiltonian
thus looses its explicit dependence on time $t$, and can be propagated
with one of the available highly accurate methods for solving the
TDSE with time-independent Hamiltonian.\cite{RonnieReview88}

Most of the vast literature on the interaction of matter with
time-dependent fields in general \cite{bandrauk91,engel98,fisher04,bandrauk06} 
and on coherent control in particular
\cite{k67,ohtsuki98,ohtsuki08,rabitz08,GollubPRA08}
ignores the effect of time 
ordering. Popular approaches include Runge-Kutta
schemes,\cite{magnus,KilinJCP09,TremblayJCP04}  
the standard Chebychev propagator with
 very small time step,\cite{MyPRA04} and the split
 propagator.\cite{feit,bandrauk91,GollubPRA08} 
Naively it is assumed that if the time step is small enough the calculation 
with an explicit time-dependent Hamiltonian can be made to converge. 
The difficulty is that this convergence is very slow -- second order in
the time step if the  
Hamiltonian is stationary in the time interval and third order if the
second order Magnus approximation is used.\cite{k79,KormannJCP08}
Additionally in many cases the error accumulates in
phase\cite{k71,k92} so that common indicators of error such as   
deviation from unitarity are misleading.

In order to obtain high quality simulations of explicitly time
dependent problems a new approach has to be developed. 
The ultimate $(t,t')$ method cannot be used in practice since it becomes
prohibitively expensive in realistic simulations. 
On the other hand we want to maintain  the exponential convergence property of
spectral decomposition such as the Chebychev propagator.
The solution is an iterative implementation of the Chebychev
propagator for inhomogeneous equations 
such that it can overcome the time ordering issue.

The paper is organized as follows. The formal solution to the problem
is introduced in Section~\ref{sec:formalsol}: 
The TDSE for an explicitly time-dependent Hamiltonian is 
rewritten as an inhomogeneous TDSE. The inhomogeneity is calculated
iteratively and converges in the limit of many iterations. At each step
of the iteration, an inhomogenenous TDSE is solved by a Chebychev
propagator which is based on a polynomial expansion of the
inhomogeneous term.\cite{NdongJCP09}
The resulting algorithm is
outlined explicitly in Section~\ref{sec:algoito} and applied to 
three different examples in Section~\ref{sec:appl1}. Its high accuracy
is demonstrated and its efficiency is discussed in comparison to other
approaches. Section~\ref{sec:concl} concludes. 


\section{Formal solution}
\label{sec:formalsol}

The Hamiltonian, $\Op{H}$, describing the interaction of a quantum system with a
time-dependent external field typically consists of a field-free,
time-independent part, $\Op{H}_0$, and an interaction term, $\Op{W}(t) = \Op{\mu}
E(t)$. 
The TDSE for such a Hamiltonian (setting $\hbar=1$),
\begin{equation}
   i\frac{\partial}{\partial t} |\psi(t)\rangle =
   \big(\Op{H}_0 +\Op{W}(t)\big)|\psi(t)\rangle \,,
   \label{eq:homo}
\end{equation}
is solved numerically by dividing the overall propagation time 
$[0,T]$ into  short time intervals  $[t_n, t_{n+1}]$,
each of length $\Delta t$. 
A two-stage approach is employed. First, the formal solution of the
TDSE is considered. The term arising from the explicit time-dependence
of the Hamiltonian is approximated iteratively. The iterative loop
thus takes care of the time ordering. Second, at each step of the
iteration, an inhomogeneous Schr\"odinger equation is obtained. It is
solved with the recently introduced Chebychev propagator for
inhomogeneous Schr\"odinger equations.\cite{NdongJCP09}

\subsection{Iterative time ordering}

The TDSE, Eq.~(\ref{eq:homo}), is rewritten to capture the 
time-dependence within the interval $[t_n, t_{n+1}]$, 
\begin{equation}
   i\frac{\partial}{\partial t} |\psi(t)\rangle = \big(\Op{H}_0
   +\Op{W}_n\big)|\psi(t)\rangle \, + \,
   \big(\Op{W}(t)-\Op{W}_n\big)|\psi(t)\rangle \,.
   \label{eq:homo2}
 \end{equation}
Here, $\Op{W}_n$ is the value of  $\Op{W}(t)$ at the
midpoint of the propagation interval,
$\Op{W}_n = \Op{W}\left(  \frac{t_{n+1}+t_n}{2} \right)$.
The formal solution of Eq.~(\ref{eq:homo2}) is given by
\begin{eqnarray}
  |\psi(t)\rangle &=& e^{-i\Op{H}_n (t-t_n)}|\psi(t_n)\rangle- \nonumber \\ 
   && i\int_{t_n}^{t}{e^{-i\Op{H}_n (t-\tau)}\Op{V}_n(\tau)|\psi(\tau)\rangle d\tau}\,,
  \label{eq:formalsol}
\end{eqnarray}
where $\Op{H}_n = \Op{H}_0 + \Op{W}_n$ denotes the  part that is
independent of time in $[t_n, t_{n+1}]$ and 
$\Op{V}_n(t) = \Op{W}(t)-\Op{W}_n$ the time-dependent part.
Eq.~(\ref{eq:formalsol}) is subjected to an iterative loop,
\begin{eqnarray}
  |\psi_k(t)\rangle &=& e^{-i\Op{H}_n (t-t_n)}|\psi_k(t_n)\rangle -\nonumber \\
 && i\int_{t_n}^{t}{e^{-i\Op{H}_n (t-\tau)}\Op{V}_n(\tau)|\psi_{k-1}(\tau)\rangle d\tau}\,,
  \label{eq:formalsolito}
\end{eqnarray}
The solution at the $k$th step of the iteration, $|\psi_k\rangle$, is calculated    
from the formal solution,  Eq.~(\ref{eq:formalsol}),  by replacing
$|\psi_k\rangle$ in the second term on the right-hand side of
Eq.~(\ref{eq:formalsol}) by $|\psi_{k-1}\rangle$ which is  known from
the previous step.

In this approach, time ordering is achieved by converging 
$|\psi_{k-1}\rangle$ to $|\psi_k\rangle$ as the iterative scheme proceeds.
This is equivalent to the derivation of the Dyson series. Starting
from the equation of motion for the time evolution
operator, 
\[
i \frac{\partial}{\partial t}\Op{U}(t,0) = \Op{H}(t)\Op{U}(t,0)\,,
\]
the formal solution for the time evolution operator,
\begin{equation}
  \label{eq:intU}
  \Op{U}(t,0) = -i \int_0^t \Op{H}(t_1) \Op{U}(t_1,0) dt_1 \,,  
\end{equation}
is iteratively inserted in the right-hand side, i.e.
\begin{eqnarray*}
  \Op{U}(t,0) &=& -i \int_0^t\int_0^{t_1} \Op{H}(t_1)
  \Op{H}(t_2)\Op{U}(t_2,0) dt_2 dt_1\,, \\
  \ldots &&\\
  \Op{U}(t,0) &=& -i \int_0^t\int_0^{t_1}\ldots\int_0^{t_{n-1}} \\ &&\Op{H}(t_1) 
  \Op{H}(t_2)\ldots \Op{H}(t_n)\Op{U}(t_n,0) dt_n \ldots dt_2 dt_1 \,,\\
\end{eqnarray*}
where $\Op{U}(t_n,0)$ goes to $\openone$ as $t_n$ becomes smaller and
smaller. 
Our formal solution, Eq.~(\ref{eq:formalsol}) is equivalent to
Eq.~(\ref{eq:intU}). An alternative approach to time ordering is given
by the Magnus expansion which is based on the group properties of
unitary time evolution.\cite{magnus} In the limit of convergence, the
Magnus and the Dyson series are completely equivalent, but low-order
approximations of the two differ.\cite{magnus} Our iterative scheme
corresponds to the limit of convergence (with respect to machine
precision).

\subsection{Equivalence to an inhomogeneous TDSE}

Differentiating  Eq.~(\ref{eq:formalsolito}) with respect to time,  
an inhomogeneous Schr\"odinger equation at each step $k$ of the
iteration is obtained,
\begin{equation}
  \frac{ \partial}{\partial t} |\psi_k(t)\rangle = 
   -i\Op{H}_n|\psi_k(t)\rangle + |\Phi_{k-1}(t)\rangle\,.
  \label{eq:inho}
\end{equation}
The inhomogeneity is given by
\begin{equation}
  \label{eq:inhomogeneity}
  |\Phi_{k-1}(t)\rangle = -i \Op{V}_n(t)|\psi_{k-1}(t)\rangle \,.
\end{equation}
Eq.~(\ref{eq:inho}) can be solved by approximating the inhomogeneous 
term globally within $[t_n, t_{n+1}]$, i.e. by expanding it into
Chebychev polynomials,
\begin{equation}
  \label{eq:Polyexpan}
  |\Phi_{k-1}(t)\rangle \approx \sum_{j=0}^{m-1}P_j(\bar t)|\bar
  \Phi_{k-1,j}\rangle  \,. 
\end{equation}
$P_{k-1,j}$ denotes the Chebychev polynomial of order $j$ with expansion coefficient
$|\bar \Phi_{k-1,j}\rangle$,  and  $\bar t = 2(t-t_n)/\Delta t - 1$
with $t \in [t_n, t_{n+1}]$ is a rescaled time.\cite{NdongJCP09}

The expansion  coefficients, $|\bar \Phi_{k-1,j}\rangle$,  in
Eq.~(\ref{eq:Polyexpan}) are given by
\begin{equation}
  \label{eq:barphi}
|\bar \Phi_{k-1,j}\rangle   = \frac{2-\delta_{j0}}{\pi} \int_{-1}^1
       \frac{  |\Phi_{k-1}(\bar t)\rangle P_j(\bar t)}{\sqrt{1-\bar t^2}} d \bar t  \,.
\end{equation}
Since $|\Phi_{k-1}(\bar t)\rangle$ is known at each point 
in the interval and in particular at the zeros, $\bar t_i$, of the $m$th Chebychev
polynomial,  the integral in Eq.~(\ref{eq:barphi}) can be rewritten
by applying a Gaussian quadrature,\cite{RoiPRA00} yielding
\begin{equation}
  \label{eq:barphisum}
    |\bar \Phi_{k-1,j}\rangle  = \frac{2-\delta_{j0}}{m}\sum_{i=0}^{m-1} 
    |\Phi_{k-1}(\bar t_i)\rangle P_j(\bar t_i) \,. 
\end{equation}
Due to the fact that the  Chebychev polynomials can be expressed in terms of
cosines, Eq.~(\ref{eq:barphisum}) is equivalent to a cosine transformation.
Thus the expansion coefficients, $|\bar \Phi_{k-1,j}\rangle$, can
easily be obtained numerically by fast cosine transformation.

The expansion into Chebychev polynomials, if converged, is equivalent
to the following alternative expansion,
\begin{equation}
  \label{eq:monic}
  \sum_{j=0}^{m-1}P_j(\bar t)|\bar \Phi_{k-1,j}\rangle =
    \sum_{j'=0}^{m-1} \frac{(t-t_n)^{j'}}{j'!}|\Phi_{k-1}^{(j')}\rangle\,.
\end{equation}
Once the coefficients of the Chebychev expansion, $|\bar
\Phi_{k-1,j}\rangle$, are known, 
the transformation described in Appendix~\ref{app:trafo} is used to generate 
the coefficients $|\Phi_{k-1}^{(j')}\rangle$ in 
Eq.~(\ref{eq:monic}).

Approximating the  inhomogeneous term by the right-hand side of
Eq.~(\ref{eq:monic}), the formal solution
of Eq.~(\ref{eq:inho}) can be written\cite{NdongJCP09}
\begin{equation}
  |\psi_k(t)\rangle = \sum_{j=0}^{m-1} \frac{(t-t_n)^j}{j!} |\lambda^{(j)}_{k-1} \rangle +
  \Op{F}_m |\lambda^{(m)}_{k-1}\rangle\,,
\label{eq:formalsolhil}
\end{equation}
where the $|\lambda^{(j)}_{k-1}\rangle$ are obtained recursively,
\begin{eqnarray}
  \label{eq:lambdas}
  |\lambda^{(0)}_{k-1} \rangle &=& |\psi (t_n) \rangle\, , \\
  |\lambda^{(j)}_{k-1}  \rangle &=& -i \Op{H}_n|\lambda^{(j-1)}_{k-1} \rangle +
  |\Phi^{(j-1)}_{k-1}  \rangle, \nonumber\\ && \quad\quad 1\leq j \leq m \,.\nonumber
\end{eqnarray}
$\Op{F}_m$ is a function of $ \Op{H}_n$ and is given by
\begin{eqnarray}
  \label{eq:Fm}
  &\Op{F}_m& =  \\
  &&(-i\Op{H}_n)^{-m} \left (e^{-i\Op{H}_n (t-t_n)} -
    \sum_{j=0}^{m-1} \frac{(-i \Op{H}_n (t-t_n))^j}{j!} \right )\,. \nonumber
\end{eqnarray}
Taking the derivative of Eq.~(\ref{eq:formalsolhil}) with respect to
time,  the inhomogeneous Schr\"odinger equation is recovered after
some algebra.\cite{NdongJCP09}

Alternatively, Eq.~(\ref{eq:inhomogeneity}) can be inserted into
Eq.~(\ref{eq:formalsolito}), replacing $|\Phi_{k-1}\rangle$ by its
polynomial approximation,  Eq.~(\ref{eq:Polyexpan}),
\begin{eqnarray}
  |\psi(t)\rangle &=& e^{-i\Op{H}_n t}|\psi(0)\rangle + \nonumber \\
  && e^{-i\Op{H}_n t} \sum_{j=0}^{m-1} 
   \int_0^t{e^{i\Op{H}_n \tau} \frac{\tau^j}{j!}|\Phi^{(j)} \rangle d\tau}
  \label{eq:formalsolproof}
\end{eqnarray}
(without any loss of generality, $t_n$ has been set to  zero).
Defining 
\begin{equation}
   \Op{\alpha}_j  =  e^{-i\Op{H}_n t}  \int_0^t{e^{i\Op{H}_n \tau}
     \frac{\tau^j}{j!}  d\tau}\,,
  \label{eq:alphaj}
\end{equation}
and integrating Eq.~(\ref{eq:alphaj}) by parts,  one obtains
\begin{eqnarray}
  \label{eq:recuralphaj}
   \Op{\alpha}_j  &=&  (-i\Op{H}_n)^{-1} \left(e^{-i\Op{H}_n t}\Op{\alpha}_{j-1} -
   \frac{t^j}{j!}\openone \right)\,,  \\
   && \quad 1\le j\le m-1 \,, \nonumber\\
   \Op{\alpha}_0  &=&  (-i\Op{H}_n)^{-1}  \left(e^{-i\Op{H}_n t} - \openone\right)\,.
  \label{eq:alpha0}
\end{eqnarray}
By induction, it follows that
\begin{equation}
   \Op{\alpha}_j  =  (-i\Op{H}_n)^{-(j+1)} \left(e^{-i\Op{H}_n t}- 
       \sum_{a=0}^{j} \frac{ (-i\Op{H}_nt)^{a}} {a!}\right)\,.
  \label{eq:Fj}
\end{equation}
Defining 
\begin{equation}
\Op{F}_{j+1} = (-i\Op{H}_n)^{-(j+1)} \left(e^{-i\Op{H}_n t}- 
              \sum_{a=0}^{j} \frac{ (-i\Op{H}_nt)^{a}} {a!}\right)\,,
\end{equation}
Eq.~(\ref{eq:formalsolproof}) becomes
\begin{equation}
  |\psi(t)\rangle = e^{-i\Op{H}_n t}|\psi(0)\rangle
  + \sum_{j=0}^{m-1} \Op{F}_{j+1} |\Phi^{(j)} \rangle \,,
  \label{eq:formalsolendproof}
\end{equation}
which was shown to be  equivalent to 
Eq.~(\ref{eq:formalsolhil}).\cite{NdongJCP09}

The algorithm for solving the TDSE with explicitly time-dependent
Hamiltonian is thus based on evaluating the integral of the formal solution,
Eq.~(\ref{eq:formalsol}), in an iterative fashion. At each step $k$ of the
iteration,  the inhomogeneous TDSE, Eq.~(\ref{eq:inho}), 
is solved by applying the propagator of Ref.~\onlinecite{NdongJCP09}
within each short time interval $[t_n,t_{n+1}]$.

Once convergence with respect to the iteration $k$ is reached, 
the inhomogeneous term becomes constant with respect to $k$.


\section{Outline of the algorithm}
\label{sec:algoito}

We assume that the action of the Hamiltonian on a wavefunction can be
efficiently computed.\cite{RonnieReview88}
Then the complete propagation time interval $[0,T]$ is split into small
time intervals, $[t_n,t_{n+1}]$. For each time step $[t_n,t_{n+1}]$, 
the implementation of the Chebychev propagator with iterative time
ordering involves an outer loop over the iterative steps $k$ for time
ordering and an inner loop over $j$ for the solution of the
(inhomogeneous) Schr\"odinger equation for each $k$.
\begin{enumerate}
\item
  Preparation:
  Set a local time grid $\{\tau_l\}$ for each short-time interval 
  $[t_n,t_{n+1}]$. In order to calculate the expansion coefficients of the
  inhomogeneous term by cosine transformation, the $N_t$ sampling
  points $\{\tau_l\}$ are chosen to be the roots of the Chebychev polynomial
  $P_{N_t}$ of order $N_t$.
  The number of sampling points, $N_t$, is not known in advance. One
  thus has to provide an initial guess and check below, in step 3.i,
  that it is equal to or larger than the number of Chebychev
  polynomials required to expand the inhomogeneous term,
  \begin{equation}
    \label{eq:Nt}
    N_t\ge m\,.  
  \end{equation}
  If $N_t$ is much larger than $m$, it is worth to decrease it (subject
  to the bound of Eq.~(\ref{eq:Nt})) and to recalculate the
  $\{\tau_l\}$. The number of propagation steps within $[t_n,t_{n+1}]$
  is then reduced to its mininum.
\item The propagation for $k=0$ solves the Schr\"odinger
  equation for the time-independent Hamiltonian $\Op{H}_n = \Op{H}_0 + \Op{W}_n$,
  \[
  i\frac{\partial}{\partial \tau} |\psi_0(\tau)\rangle = \big(\Op{H}_0
  +\Op{W}_n\big)|\psi_0(\tau)\rangle\,,
  \]
  with initial condition $|\psi_0(t=t_n)\rangle  = |\psi(t_n)\rangle$.
  A standard Chebychev propagator is employed to this end.
  Note that for $k=0$, the same time grid $\{\tau_l\}$ needs to be used as for
  $k>0$ because the inhomogeneous term for $k=1$ is calculated from
  the zeroth order solution,  $|\psi_0(t)\rangle$.
  Since the $\{\tau_l\}$ are not equidistant, the Chebychev expansion
  coefficients of the standard propagator,
  $e^{-i\Op{H}_n\Delta \tau_l}$,  need to be calculated for each time step within
  $[t_n,t_{n+1}]$, where $\Delta \tau_l=\tau_{l+1}- \tau_l$,
  $l=1,N_t-1$.
\item The $k>0$ propagation solves an inhomogeneous Schr\"odinger
  equation, cf.    Eq.~(\ref{eq:inho}), with the initial condition
  $|\psi_k(t=t_n)\rangle=|\psi_0(t=t_n)\rangle=|\psi(t_n)\rangle$. This
  is achieved by the Chebychev propagator for inhomogeneous
  Schr\"odinger equations,\cite{NdongJCP09}
  i.e. Eq.~(\ref{eq:formalsolhil}), and involves the following steps: 
  \begin{enumerate}\renewcommand{\theenumii}{\roman{enumii}}
  \item Evaluate the inhomogeneous term,
    $|\Phi_{k-1}(\tau) \rangle = -i
    \big(\Op{W}(\tau)-\Op{W}_n\big)|\psi_{k-1}(\tau)\rangle$.
  \item Calculate the expansion coefficients of the inhomogeneous term,
    cf. Eqs.~(\ref{eq:Polyexpan}) and (\ref{eq:monic}). The 
    Chebychev expansion coefficients $|\bar\Phi_{k-1,j}\rangle$ are obtained by
    cosine transformation of $|\Phi_{k-1}(\tau) \rangle$.\cite{NdongJCP09}
    The coefficients $|\Phi^{(j)}_{k-1}\rangle$ are evaluated 
    from the Chebychev expansion coefficients $|\bar\Phi_{k-1,j}\rangle$
    using the recursive relation given in
    Eqs.~(\ref{eq:barphitophiend1}) and (\ref{eq:barphitophiend2}). The
    order $m$ of   the expansion is chosen such that ratio of the
    smallest to the largest Chebychev coefficient     becomes smaller
    than the specified error $\epsilon$, 
    \begin{equation}
      \label{eq:Chebylimit}
      \frac{\|\bar\Phi_{k-1,m+1}\|}{\|\bar\Phi_{k-1,0}\|} <
      \epsilon \,.
    \end{equation}
    To obtain high accuracy,
    $\epsilon$ may correspond to the machine precision.
    \footnote{If the Chebychev coefficients can be calculated based on
      an analytical expression, the smallest Chebychev coefficient
      itself can be pushed below machine
      precision. This is the case, for example, for
      the standard Chebychev propagator where the expansion
      coefficients of the function $e^{-ix}$ are given in terms of
      Bessel functions. Here, our accuracy is limited to the relative
      error specified by 
      Eq.~(\ref{eq:Chebylimit}) because the expansion coefficients can only
      be obtained numerically by fast cosine transformation.
      }
  \item Calculate the Chebychev expansion coefficients of  $\Op{F}_m$, cf.
    Eq.~(\ref{eq:Fm}), also by cosine transformation. The number of
    terms in this Chebychev expansion is also determined by the
    relative magnitude of the coefficients,
    analogously to Eq.~(\ref{eq:Chebylimit}).
  \item Determine all $|\lambda^{(j)}_{k-1}\rangle$ required in 
    Eq.~(\ref{eq:formalsolhil}) by evaluating Eq.~(\ref{eq:lambdas}).
  \item Construct the solution  $|\psi_k(t=t_{n+1})\rangle$ according to
    Eq.~(\ref{eq:formalsolhil}). 
  \end{enumerate}
\item Convergence is reached when 
  $|\psi_{k-1}(t_{n+1})\rangle$ and $|\psi_{k}(t_{n+1})\rangle$ become
  indistinguishable, 
  \[
  \|\psi^{k-1}(t_{n+1}) - |\psi^{k}(t_{n+1})\|
  < \epsilon\,,
  \]
  and the desired solution of the Schr\"odinger equation with explicitly
  time-dependent Hamiltonian is obtained,
  $|\psi(t_{n+1})\rangle = |\psi_{k}(t_{n+1})\rangle$.
\end{enumerate}

The only parameter of the algorithm is the pre-specified error
$\epsilon$. It determines the number of iterative terms $k$ and the order
of the inhomogeneous propagator $m$. Furthermore, to execute the
algorithm, the user has to provide, besides $\epsilon$, an initial
guess for the number of sampling points of the local time grid, $N_t$.


\section{Examples}
\label{sec:appl1}

We test the accuracy and efficiency of the algorithm for three
examples of increasing complexity. The first two examples, a driven
two-level atom and a linearly driven harmonic oscillator, are
analytically solvable. We can therefore compare the numerical to the
analytical solution and establish the accuracy of the Chebychev
propagator with iterative time ordering. 
For the third example, wave packet interferometry
in two oscillators coupled by a field, no analytical solution is
known. The Chebychev propagator with iterative time ordering thus
serves as a reference solution to which less accurate methods can be
compared. 

\subsection{Driven two-level atom}
\label{subsec:example1} 

The Hamiltonian for a two-level atom driven resonantly by a laser
field in the rotating-wave approximation reads\cite{AllenEberly}
\begin{equation}
\Op{H} =
\begin{pmatrix}
  0 & \Op{\mu}E(t) \\
  \Op{\mu}E(t) & 0 \\
\end{pmatrix}  \,,
\end{equation}
where the field is of the form
\begin{equation}
E(t) = \frac{1}{2}E_0 S(t)\,,
\end{equation}
and $S(t)$ denotes the envelope of the field.
We take the strength of the transition dipole to be $\mu=1\,$a.u., the
final propagation time $T = 9000\,$a.u.,  and the shape function
\begin{equation}
   S(t) = \sin^2\left(\frac{\pi t}{T}\right)\,.
   \label{eq:shapefunction}
\end{equation}
Analytically, the time evolution of the amplitudes,
\begin{equation}
  |\psi(t)\rangle =
  \begin{pmatrix}
    c_g(t) \\
    c_e(t)
  \end{pmatrix}\,,
\end{equation}
is obtained as
\begin{eqnarray}
  c_g^{\mathrm{ana}}(t) &=&  \cos\left[\frac{1}{4}\mu
    E_0\left(t-\frac{T}{2\pi}
      \sin\left(\frac{2\pi t}{T}\right)\right) \right]\,,\\
  c_e^{\mathrm{ana}}(t) &=& i \sin\left[\frac{1}{4}\mu E_0\left(t-\frac{T}{2\pi}
    \sin\left(\frac{2\pi t}{T}\right)\right) \right]\,.
\end{eqnarray}
Initially the two-level system is assumed to be in the ground state, $c_g(t=0)=1$,
$c_e(t=0)=0$. The pulse amplitude is chosen
to yield a $\pi$-pulse, such that
$c_g^{\mathrm{ana}}(t=T)=0$, $c_e^{\mathrm{ana}}(t=T)=1$.  

\begin{figure}[tb]
  \centering
  \includegraphics[width = 0.95\linewidth]{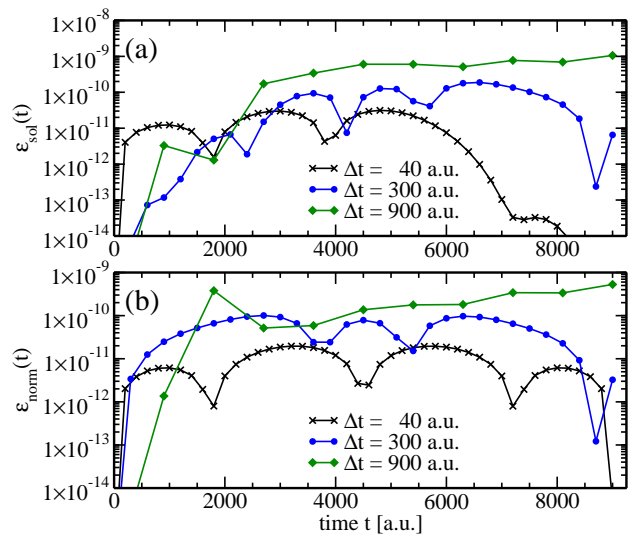}
  \caption{(color online) Strongly driven two-level atom propagated
    with iteratively time ordering Chebychev propagator:
    error of the solution, $\varepsilon_\mathrm{sol}(t)$ (a) and
    deviation of the norm of $|\psi(t)\rangle$ from unity,
    $\varepsilon_\mathrm{norm}(t)$ (b).
  }
  \label{fig:errtwolevel}
\end{figure}

Defining at each time step the errors,
\begin{equation}
  \mathrm{\varepsilon_{sol}}(t)  = \left | |c_g^{\mathrm{ana}}(t)|^2 - 
    |c_g(t)|^2\right|\,, 
\end{equation}
and 
\begin{equation}
  \mathrm{\varepsilon_{norm}}(t) = \big|1-  \langle \psi(t)|\psi(t)\rangle \big| \,, 
\end{equation}
we measure the deviation  of the numerical from the analytical solution 
and the deviation of the norm of $|\psi(t)\rangle$ from unity.
The time evolution of $\mathrm{\varepsilon_{sol}}(t)$
and $\mathrm{\varepsilon_{norm}}(t)$ is shown in
Fig.~\ref{fig:errtwolevel} for the Chebychev
propagator with iterative time ordering. 
The maximum errors occuring during the propagation,
$\mathrm{\varepsilon_{norm}^{\max}}$ and
$\mathrm{\varepsilon_{sol}^{\max}}$,
are also summarized in Table \ref{tab:errtwolevel}.
\begin{table}[tb]
  \centering
  \begin{tabular}{|c| c| c| c| c| c| c| c|}\hline
    $\Delta t$& $N_t$ & $ m_k$ &  $ N_\mathrm{Cheby}$ &
    $\mathrm{\varepsilon_{sol}^{\max}}$  &
    $\mathrm{\varepsilon_{norm}^{\max}}$  & CPU time  &
    $k_{\max}$ \\ \hline  
    10 & 6   & $4\,$    &      10                 &
    $1.7\cdot 10^{-11}$   &   $ 1.1\cdot 10^{-11}$   &           $23\,$s   & 3
    \\
    & 12  & $4\,$      &          10              &  
    $1.7\cdot 10^{-11}$   &   $ 1.1\cdot 10^{-11}$   &           $ 47\,$s    & 3 \\
    \hline
    20 & 7   & $5\,$     &         11                 &
    $4.1\cdot 10^{-11}$   &   $ 1.8\cdot 10^{-11}$   &           $14\,$s   & 4
    \\
    &14  &$5\,$       &          11               &
    $4.1 \cdot10^{-11}$        &   $1.8\cdot 10^{-11}$ &        $ 29\,$s & 4\\
    \hline
    40 & 7   & $5\,$     &       14                 &
    $3.1\cdot 10^{-11}$   &   $ 1.2\cdot 10^{-11}$   &           $8\,$s   & 4
    \\
    &14  &$5\,$       &          14               &
    $3.1 \cdot10^{-11}$        &   $1.2\cdot 10^{-11}$ &        $ 15\,$s & 4\\
    \hline
    80 &  8  &  6   &          16            & 
    $1.9\cdot10^{-11}$        &    $1.1\cdot10^{-11}$         & $6\,$s     &   5  \\
     &  16  &  6   &          16              & 
     $1.9\cdot10^{-11}$      &  $1.1\cdot10^{-11}$  &         $11\,$s     &   5   \\
      \hline
    100 &  9  &  7   &          17            & 
    $8.3\cdot10^{-11}$        &    $4.0\cdot10^{-11}$         & $5\,$s     &   5  \\
     &  18  &  7   &          17              & 
     $8.3\cdot10^{-11}$      &  $4.0\cdot10^{-11}$  &         $9\,$s     &   5   \\
    \hline
    300 &  10  &  8   &          29          & 
    $1.9\cdot10^{-10}$        &    $1.0\cdot10^{-10}$         & $3.4\,$s     &   6  \\
     &  20  &  9   &          29              & 
    $1.9\cdot10^{-10}$        &    $1.0\cdot10^{-10}$         & $5.3\,$s     &   6  \\
    \hline
    600 &  12  &  10   &          32            & 
    $5.7\cdot10^{-10}$        &    $3.1\cdot10^{-10}$         & $2.6\,$s     &   6  \\
     &  24  &  10   &          32              & 
    $5.7\cdot10^{-10}$        &    $3.1\cdot10^{-10}$         & $4.2\,$s     &   6  \\
    \hline
    700 &  12  &  10   &          33            & 
    $7.8\cdot10^{-10}$        &    $3.6\cdot10^{-10}$         & $2.1\,$s     &   7  \\
     &  24  &  10   &          33              & 
    $7.8\cdot10^{-10}$        &    $3.6\cdot10^{-10}$         & $3.8\,$s     &   7  \\
    \hline
    800 &  14  &  12   &         35           & 
    $5.2\cdot10^{-10}$        &    $2.3\cdot10^{-10}$         & $2.5\,$s     &   8  \\
     &  28  &  12   &          35              & 
     $5.2\cdot10^{-10}$      &  $2.3\cdot10^{-10}$  &         $4.3\,$s     &   8   \\
    \hline
    900 &  15  &  13   &          36           & 
    $1.1\cdot10^{-9}$        &    $5.3\cdot10^{-10}$         & $3.0\,$s     &   8  \\
     &  30  &  13   &          36              & 
    $1.1\cdot10^{-9}$        &    $5.3\cdot10^{-10}$         & $5.1\,$s     &   8  \\
    \hline
    1000 &  17    &     15        &         38            & 
    $3.6\cdot10^{-9}$        &    $7.0\cdot10^{-10}$         & $3.5\,$s     &   9  \\
     &  34  &  15   &          38              & 
    $3.6\cdot10^{-9}$        &    $7.0\cdot10^{-10}$         & $5.8\,$s     &   9  \\
    \hline
  \end{tabular}
  \caption{ 
    The maximum error of the solution,
    $\mathrm{\varepsilon_{sol}^{\max}}$,
    and the maximum deviation of the norm from unity,
    $\mathrm{\varepsilon_{norm}^{\max}}$,
    occuring in the overall propagation time are listed together with
    the required CPU time
    for several short time intervals $\Delta t$.
    $N_t$ denotes the number of sampling points within $\Delta t$,
    $N_\mathrm{Cheby}$ the largest number of Chebychev coefficients in the
    expansion of $\Op{F}_m$,
    $m_k$ the order of the  expansion of the inhomogeneous
    term and $k_{\max}$ the largest number of the iterations for
    time ordering occuring for all time intervals $[t_n,t_{n+1}]$.
  }
  \label{tab:errtwolevel}
\end{table}
For time steps, $\Delta t=t_{n+1}-t_n$, up to about $T/100$, 
the maximum errors occuring during the propagation, 
$\mathrm{\varepsilon_{sol}^{\max}}$, are of the order of $10^{-11}$.  
If the time step is further increased to about $T/10$, the maximum
errors are of the order of $10^{-9}$. The increase in
$\mathrm{\varepsilon_{sol}^{\max}}$ is accompanied by an increase in
$\mathrm{\varepsilon_{norm}^{\max}}$ as the time steps become larger,
cf. Table~\ref{tab:errtwolevel}. 
The deviation  from unitarity indicates that the
error is due to the Chebychev expansion of the time evolution which
becomes unitary only once the series is converged. The limiting factor
here is the accuracy of the numerically obtained
Chebychev expansion coefficients. This effect becomes more severe, as
the argument of the Chebychev polynomials, $\Delta t\Delta E$ (and thus
the largest expansion coefficient) becomes larger and larger.

The errors obtained by the Chebychev propagator with iterative time
ordering of the order of $10^{-11}$ to $10^{-9}$ have to be compared
to those obtained by the standard Chebychev propagator, i.e. neglecting all
effects due to time ordering. The latter yields maximum solution errors,
$\mathrm{\varepsilon_{sol}^{\max}}$, of the order of $10^{-4}$ for
$\Delta t = 10\,$a.u.$=T/900$ and $10^{-3}$ for $\Delta t = 40\,$a.u. 
The smallest $\mathrm{\varepsilon_{sol}^{\max}}$ that can be achieved
without time ordering is of the order of $10^{-6}$ for $\Delta t =
10^{-2}\,$a.u.$=T/900000$. 
Thus the numerical results obtained with the iterative method are highly
accurate compared to those obtained by the standard Chebychev
propagator neglecting time ordering.

Regarding the numerical efficiency of the Chebychev propagator with
iterative time ordering, several conclusions can be drawn from 
Table~\ref{tab:errtwolevel}. First of all, it is absolutely sufficient
to choose the number of sampling points within the interval $\Delta
t$, $N_t$, 
only slightly larger than the order of the expansion of the
inhomogeneous term, $m_k$. Doubling $N_t$ doesn't yield better
accuracy but requires more CPU time. Second, we expect an optimum in
terms of CPU time as $\Delta t$ is increased. A Chebychev expansion
always comes with an offset and becomes more efficient as more terms in the
expansion but less time steps are required (this concerns both
Chebychev expansions, that for the inhomogeneous term of order $m_k$
and that for the time evolution operator, i.e. for the $\Op{F}_m$, of order
$N_\mathrm{Cheby}$). However, this trend
is countered by a higher number of iterations for time
ordering, $k_\mathrm{max}$.
According to Table~\ref{tab:errtwolevel}, the optimum in
terms of CPU time is found for $\Delta t \approx 700\,$a.u. Finally,
the order required in the Chebychev expansion of the inhomogeneous
term, $m_k$, stays comparatively small, well below the values where
the transformation between the Chebychev coefficients and the
polynomial coefficients becomes numerically instable,
cf. Appendix~\ref{app:trafo}.

\subsection{Driven harmonic oscillator}
\label{subsec:example2} 

As a second example, we consider a harmonic oscillator of mass $m=1\,$a.u.
and frequency $\omega=1\,$a.u.  driven by a linearly polarized field. The 
time-dependent Hamiltonian is given by
\begin{equation}
  \Op{H}(r;t) = -\frac{1}{2}\frac{\partial^2}{\partial r^2}+\frac{1}{2}r^2 +
                 r E_0 S(t) \cos(\omega_0t) \,,
  \label{eq:hamoscill}
\end{equation}
where $E_0$ is the maximum field amplitude, $S(t)$ the shape
function given by Eq.~(\ref{eq:shapefunction}), 
and $\omega_0$ is the frequency of the driving field. The final time is set to
$T = 100\,$a.u. The Hamiltonian is represented on a Fourier
grid\cite{RonnieReview88}
with $N_{\mathrm{grid}} =128$ grid points, and
$r_\mathrm{max} = 10\,$a.u.$=-r_\mathrm{min}$.
The transition probabilities and expectation values of
position and momentum as a function of time are known
analytically.\cite{Ronniettprime,SalaminJPM95}

Taking the initial wave function $|\psi(t=0)\rangle$ to be the ground state of
the harmonic oscillator, we again measure the deviation of the
numerical  from the analytical solution, $\mathrm{\varepsilon_{sol}}$,
and the deviation of the norm of
$|\psi(t)\rangle$ from unity,  $\mathrm{\varepsilon_{norm}}$, for the
time-dependent probability of the oscillator to be in the ground state.
The pulse amplitude is chosen to completely deplete the population
of the ground state.  

Two cases are analyzed which both correspond to strong resonant driving of
the oscillator. In the first case the
rotating-wave approximation is invoked, i.e. we set $\omega_0 = 0$. This 
eliminates the highly oscillatory term from the field, keeping only the
time-dependence of the shape function (moderate time-dependence).
In the second case, the rotating-wave approximation is avoided,
$\omega_0 = \omega$, i.e. the time-dependence of the Hamiltonian is
much stronger than in the first case (strong time-dependence).

In order to compare the Chebychev propagators with and without
time ordering, we first list the smallest
$\mathrm{\varepsilon_{sol}^{\max}}$ and $\mathrm{\varepsilon_{norm}^{\max}} $
achieved by the standard Chebychev
propagator without time ordering in Table~\ref{tab:errstandard}.
\begin{table*}[tbp]
  \centering
  \begin{tabular}{|c |c| c| c| c| c| c |}\hline
    $\Delta t$  &  $ N_\mathrm{Cheby}$ &
    $\mathrm{\varepsilon_{norm}^{\max}}$  ($\omega_0=\omega$) & $\mathrm{
    \varepsilon_{norm}^{\max}}$  ($\omega_0=0$)
    &   $\mathrm{\varepsilon_{sol}^{\max}}$  (
    $\omega_0=\omega$) &  $\mathrm{\varepsilon_{sol}^{\max}}$
    ($\omega_0=0$) &    CPU time
    \\ 
    \hline
    $ 10^{-6}\,$a.u. &     4                 &
    $6.8\cdot 10^{-9}$  &  $6.7\cdot 10^{-9}$   &   
    $ 4.6\cdot 10^{-8}$ & $6.7\cdot 10^{-9}$   &          $1\,$ h   $16\,$ m $ 28\,$s 
    \\
    $ 10^{-5}\,$a.u.  &     5               &
    $6.6\cdot 10^{-10}$  & $6.6\cdot 10^{-10}$   &   
    $ 4.7\cdot 10^{-7}$  & $4.4\cdot 10^{-9}$   &             $9\,$ m $ 26\,$s 
    \\
    $ 10^{-4}\,$a.u.  &     7                &
    $1.3\cdot 10^{-11}$  &  $ 4.7\cdot 10^{-11}$   &   
    $ 4.7\cdot 10^{-6}$  &  $4.2\cdot 10^{-8}$   &             $1\,$ m $ 28\,$s
    \\
    $ 10^{-3}\,$a.u.  &     10                &
    $6.5\cdot 10^{-12}$ & $7.3\cdot 10^{-12}$   &   
    $ 4.7\cdot 10^{-5}$ & $ 4.2\cdot 10^{-7}$   &              $ 12\,$s \\
    \hline
  \end{tabular}
  \centering
  \caption{
    Driven harmonic oscillator with ($\omega_0=0$) and without
    ($\omega_0=\omega$) the rotating-wave approximation for the
    standard Chebychev propagator without time ordering.
    $N_\mathrm{Cheby}$ is
    the  number of Chebychev polynomials required for the expansion of
    $e^{-i\Op{H}\Delta t}$.
  }
  \label{tab:errstandard}
\end{table*}
The standard Chebychev propagator was developed for time-independent
problems and is most efficient for large time steps. Here, however,
extremely small time steps $\Delta t$ have to be employed to minimize
the error due to the time-dependence of the Hamiltonian. Consequently,
the required CPU times become quickly very large. Note 
that the deviation of the norm from unity is much smaller than the error of
the solution. This means that norm conservation cannot serve as an
indicator for the error due to the time-dependence of the
Hamiltonian. This error is clearly non-negligible even for the very small
time steps shown in Table~\ref{tab:errstandard}.

Tables~\ref{tab:errharmsaphe} and \ref{tab:errharmcossaphe} compare 
the results  for the driven harmonic oscillator
obtained by the Chebychev propagator with iterative
time ordering (ITO) and without  time ordering (standard Chebychev
propagator). The rotating-wave approximation is invoked in
Table~\ref{tab:errharmsaphe}, $\omega_0=0$, and avoided in
Table~\ref{tab:errharmcossaphe}, $\omega_0=\omega$.
\begin{table*}[tbp]
  \centering
  \begin{tabular}{|c|c|c|c|c|c|c|c||c|c|c|c|c|}\hline
    \multicolumn{8}{|c||}{with iterative time ordering (ITO)} &
    \multicolumn{5}{c|}{without time ordering (standard)} 
    \\ \hline
    $\Delta t$& $N_t$ & $ m_k$ &  $ N_\mathrm{Cheby}$ &
    $\mathrm{\varepsilon_{norm}^{\max}}$ &
    $\mathrm{\varepsilon_{sol}^{\max}}$  &
    CPU time  & $k_{\max}$  &   $\Delta t$
    &  $ N_\mathrm{Cheby}$ &
    $\mathrm{\varepsilon_{norm}^{\max}}$  &  $\mathrm{\varepsilon_{sol}^{\max}}$  &
    CPU time   \\ \hline 
    0.01$\,$a.u. & 10   & $8\,$    &      9                 &
    $5.3\cdot 10^{-13} $   &   $ 5.3\cdot 10^{-13}$   &           $1\,$ m $ 54\,$s   & 2
    &0.01$\,$a.u. &     18                 &
    $1.4\cdot 10^{-12}$   &   $ 4.2\cdot 10^{-6}$   &            $ 2\,$s 
    \\
    0.02$\,$a.u. & 10   & $8\,$     &        10                 &
    $1.4\cdot 10^{-13}$   &   $ 1.4\cdot 10^{-13}$   &            $ 58\,$s    & 2
    &0.02$\,$a.u. &     18                 &
    $5.7\cdot 10^{-12}$   &   $ 8.5\cdot 10^{-6}$   &            $ 1.28\,$s 
    \\
    0.04$\,$a.u. & 10   & $8\,$     &     15                &
    $5.5\cdot 10^{-13}$   &   $ 4.5\cdot 10^{-13}$   &             $ 31\,$s    & 2
    &0.04$\,$a.u. &     29                &
    $6.15\cdot 10^{-12}$   &   $ 1.7\cdot 10^{-5}$   &            $ 1\,$s 
    \\
    0.1$\,$a.u. & 10   & $8\,$     &     24                &
    $1.4\cdot 10^{-12}$   &   $ 1.4\cdot 10^{-12}$   &            $ 27\,$s    & 3
    & 0.1$\,$a.u. &     44                &
    $1.5\cdot 10^{-12}$   &   $ 4.2\cdot 10^{-5}$   &            $ 0.52\,$s 
    \\
    \hline
  \end{tabular}
  \caption{Comparison of the Chebychev propagator with  iterative (ITO) and without
    time ordering for the 
    driven harmonic oscillator in the rotating-wave approximation ($\omega_0=0$).
    Notation as in Table~\ref{tab:errtwolevel}.
  }
  \label{tab:errharmsaphe}
\end{table*}
\begin{table*}[tbp]
  \centering
  \begin{tabular}{|c| c| c| c| c| c| c| c||c| c| c| c| c|}\hline
    \multicolumn{8}{|c||}{with iterative time ordering (ITO)} &
    \multicolumn{5}{c|}{without time ordering (standard)} 
    \\ \hline
    $\Delta t$& $N_t$ & $ m_k$ &  $ N_\mathrm{Cheby}$ &
    $\mathrm{\varepsilon_{sol}^{\max}}$  &
    $\mathrm{\varepsilon_{norm}^{\max}}$   &        CPU time  &
    $k_{\max}$ 
    & $\Delta t$  &  $ N_\mathrm{Cheby}$ &
    $\mathrm{\varepsilon_{norm}^{\max}}$   &
    $\mathrm{\varepsilon_{sol}^{\max}}$   &    CPU time 
    \\ \hline 
    0.01$\,$a.u. & 10   & $8\,$    &      9                 &
    $8.2\cdot 10^{-14}$   &   $ 3.5\cdot 10^{-13}$   &           $2\,$ m $ 5\,$s   & 3
    & 0.01$\,$a.u. &     18                 &
    $1.5\cdot 10^{-12}$   &   $ 4.6\cdot 10^{-4}$   &            $ 2\,$s 
    \\
    0.02$\,$a.u. & 10   & $8\,$     &        10                 &
    $2.5\cdot 10^{-13}$   &   $ 3.6\cdot 10^{-12}$   &           $1\,$ m $ 34\,$s    & 3
    &0.02$\,$a.u. &     23                &
    $4.7\cdot 10^{-12}$   &   $ 9.3\cdot 10^{-4}$   &            $ 1.28\,$s
    \\
    0.04$\,$a.u. & 10   & $8\,$     &     15                &
    $3.6\cdot 10^{-13}$   &   $ 5.5\cdot 10^{-11}$   &             $ 56\,$s    & 3
    &0.04$\,$a.u. &     29                &
    $8.9\cdot 10^{-12}$   &   $ 1.8\cdot 10^{-3}$   &            $ 1\,$s 
    \\
    \hline
  \end{tabular}
  \caption{Comparison of the Chebychev propagator with iterative (ITO) and without
    time ordering for the 
    driven harmonic oscillator without the rotating-wave approximation
    ($\omega_0=\omega$). 
    Notation as in Table~\ref{tab:errtwolevel}.
  }
  \label{tab:errharmcossaphe}
\end{table*}

In the case of the rotating-wave approximation, both propagators
conserve the norm on the order of $10^{-12}$. However, only the
propagator with iterative time ordering achieves an accuracy of the
solution of the same order of magnitude while the standard propagator
yields errors of the order of  $10^{-6}$ for the time steps listed in
Table~\ref{tab:errharmsaphe}.  
The smallest maximum error of the solution achieved by the standard
Chebychev propagator for $\omega_0=0$ is of the order of $10^{-8}$ 
for a norm deviation of the order of  $10^{-11}$,
cf. Tab.~\ref{tab:errstandard}. However, this requires a prohibitively
small time step, $\Delta t = 10^{-4}$ a.u.

Even for a very strongly time-dependent Hamiltonian, when
the rotating-wave approximation is not invoked  ($\omega_0 =\omega$), 
the Chebychev propagator with iterative time ordering yields similarly
accurate results, with errors of the order of $10^{-13}$,
cf. Table~\ref{tab:errharmcossaphe}. For comparison, the 
error obtained for the standard Chebychev propagator without
time ordering is of the order of $10^{-3}$ for the time steps reported
in Table~\ref{tab:errharmcossaphe}. The smallest errors 
achieved with the standard Chebychev
propagator are of the order of $10^{-6}$ for a norm 
deviation of the order of  $10^{-11}$
for extremely small time steps, cf. Table~\ref{tab:errstandard}.

\begin{figure}[btp]
  \centering
  \includegraphics[width = 0.95\linewidth]{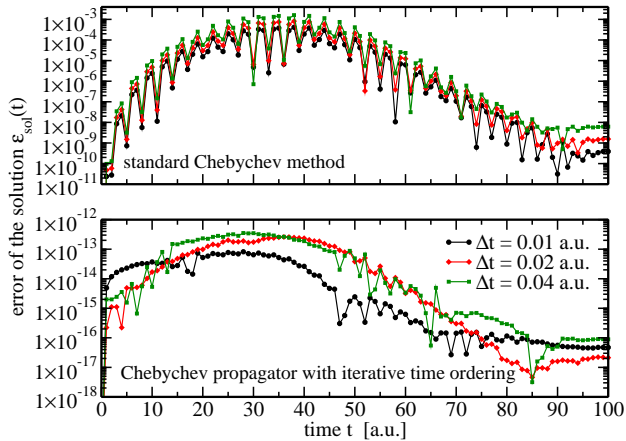}
  \caption{
  (color online) Strongly time-dependent Hamiltonian ($\omega_0 =
  \omega$): 
  Comparison of the Chebychev propagators without time ordering
  (standard) and  with iterative time ordering in terms of the
  difference between the numerical and analytical solution,
  $\mathrm{\varepsilon_{sol}(t)}$. 
  }
  \label{fig:errharmo}
\end{figure}

The error of the solution, $\mathrm{\varepsilon_{sol}}(t)$, is shown in Fig.
\ref{fig:errharmo} as a function of time for different
time steps and a very strongly time-dependent Hamiltonian, $\omega_0 = \omega$.
This illustrates the superiority of the Chebychev propagator with
iterative time ordering in terms of accuracy. A comparison of
Tables~\ref{tab:errstandard} and \ref{tab:errharmcossaphe} reveals
furthermore that the Chebychev propagator with
iterative time ordering is also more efficient than a standard
Chebychev propagator with very small time step if a high accuracy of
the solution is desired.

Since we have established the Chebychev propagator with iterative
time ordering as a highly accurate method for the solution of the TDSE
with explicitly time-dependent Hamiltonian, it is worthwhile to
compare it to alternative propagation methods for this class of
problems. In the following we will consider the 
$(t,t')$ method\cite{Ronniettprime} and a fourth-order Runge-Kutta
scheme. The $(t,t')$ method provides a numerically exact propagation
scheme by  translating the problem of time ordering into an
additional degree of freedom of a time-independent
Hamiltonian.\cite{Ronniettprime} The TDSE for the Hamiltonian in the
extended space is solved by numerically exact propagation
schemes such as the Chebychev or Newton propagators.\cite{RonnieReview94}
The $(t,t')$ method is, however, relatively rarely used in the
literature due to its 
numerical costs in terms of both CPU time and required storage.
On the other hand, Runge-Kutta schemes are extremely popular in the
literature.\cite{UlrichBenchmark00} They are potentially very
accurate if a high-order variant is employed. Note that high order
of a Runge-Kutta method implies evaluation of the Hamiltonian at
several points within the time step $[t_n,t_{n+1}]$. 

In order to achieve a fair comparison between the Chebychev propagator
with iterative time ordering and the $(t,t')$ method, first 
the parameters which yield an optimal performance of the
$(t,t')$ method for our example have to be determined. The 
required CPU time and the errors, $\mathrm{\varepsilon_{sol}^{\max}}$ and
$\mathrm{\varepsilon_{norm}^{\max}}$, as a function of the  
number of grid points, $N_{t'}$ and $N_{t}$,
are listed in Table \ref{tab:errttprime}.
\begin{table}[tbp]
  \centering
  \begin{tabular}{|c|c|c|c|c|c|}\hline
    $N_{t'}$ & $N_t $ &  $ N_\mathrm{Cheby}$ &
    $\mathrm{\varepsilon_{sol}^{\max}}$   &
    $\mathrm{\varepsilon_{norm}^{\max}}$  &        CPU time  \\
    \hline  
    1024 & 1024     &   43                 &
    $2.3\cdot 10^{-12}$   &   $ 1.2\cdot 10^{-12}$   &           $16\,$ m $ 37\,$s  
    \\
    128 & 128    &   150                 &
    $3.0\cdot 10^{-13}$   &   $ 1.3\cdot 10^{-13}$   &            $ 43\,$s   \\
    128 & 16    &      906              &
    $3.4\cdot 10^{-13}$   &   $ 1.7\cdot 10^{-13}$   &            $ 32\,$s   \\
    128 & 8    &         1740           &
    $2.4\cdot 10^{-13}$   &   $ 1.2\cdot 10^{-13}$   &            $ 30\,$s   \\
    \hline
    1024 & 1024      &   43                 &
    $ 2.5\cdot 10^{-10}$  & $3.2\cdot 10^{-7}$   &           $17\,$ m $ 43\,$s  
    \\
    2048    &   2048      &   33                 &
    $3.3\cdot 10^{-11}$   &   $ 2.8\cdot 10^{-11}$   &          $1\,$
    h  $08\,$ m $ 46\,$s \\  
    2048    &   512      &   73                 &
    $3.3\cdot 10^{-11}$   &   $ 1.6\cdot 10^{-11}$   &          $36\,$ m $ 58\,$s \\ 
    2048    &   256      &   119                 &
    $3.3\cdot 10^{-11}$   &   $ 1.6\cdot 10^{-11}$   &          $30\,$ m $ 36\,$s
    \\
    2048    &   128      &   204                 &
    $3.3\cdot 10^{-11}$   &   $ 1.6\cdot 10^{-11}$   &          $25\,$ m $ 42\,$s
    \\
    2048    &   64      &   366                 &
    $3.3\cdot 10^{-11}$   &   $ 1.6\cdot 10^{-11}$   &          $22\,$ m $ 58\,$s
    \\
    2048    &   32      &   680                 &
    $3.3\cdot 10^{-11}$   &   $ 1.6\cdot 10^{-11}$   &          $21\,$ m $ 21\,$s
    \\
    2048    &   16      &   1295                 &
    $3.3\cdot 10^{-11}$   &   $ 1.6\cdot 10^{-11}$   &          $21\,$ m $ 14\,$s
    \\
    \hline
  \end{tabular}
  \caption{Performance of the $(t,t')$ method.
    The required CPU time and the errors,
    $\mathrm{\varepsilon_{norm}^{\max}}$ and
    $\mathrm{\varepsilon_{sol}^{\max}}$, are listed for different
    numbers of sampling points of the $t'$ coordinate, $N_{t'}$  and
    different numbers of sampling points within $[0,T]$,
    $N_{t}$ together with the number of required terms in the
    Chebychev expansion, $N_\mathrm{Cheby}$.
    The upper (lower) part corresponds to $\omega_0 = 0$ ($\omega_0 = \omega$). 
  }
  \label{tab:errttprime}
\end{table}
In case of a moderate time-dependence of the Hamiltonian
corresponding to the rotating-wave approximation ($\omega=0$), a
fairly small number of grid points in both $t$ and $t'$ is
sufficient. Note that the number of points in the auxiliary
coordinate, $N_{t'}$, is not known a priori.

For a strong time-dependence, i.e. $\omega_0=\omega$,
a fairly large number of points for the auxiliary coordinate, $t'$, is
required, $N_{t'}=2048$. However, since the actual propagation
involves a time-independent Hamiltonian, large time steps can be taken
for the Chebychev propagator, resulting in the most efficient solution
when $N_t$ is small, $N_t=16$, and correspondingly the number of
Chebychev terms, $N_{cheby}$, is large. 

Table \ref{tab:erritottprime} reports the comparison between the
Chebychev propagator with iterative time ordering (ITO), the $(t,t')$
method, and the fourth-order Runge-Kutta scheme (RK4) for the
resonantly driven harmonic oscillator.
\begin{table}[tbp]
  \begin{tabular}{|c|c|c|c|c|c|c|c|}\hline
   & & $\mathrm{\varepsilon_{sol}^{\max}}$
    &  $\mathrm{\varepsilon_{norm}^{\max}}$ & CPU time\\
     \hline
    & ITO  & $5.5\cdot 10^{-13}$ &  $ 4.5\cdot 10^{-13}$ &  $ 31\,$s 
             \\
   $\omega_0=0$ & $(t,t')$ & $2.9\cdot 10^{-13}$ & $ 1.2\cdot 10^{-13}$
   &    $30\,$s \\
   & RK4  & $8.6\cdot 10^{-10}$ & $ 3.5\cdot 10^{-13}$ &  $38\,$ m $ 24\,$s \\
    \hline
   &  ITO  &  $2.5\cdot 10^{-13}$  & $3.6\cdot 10^{-12}$ &  $1\,$ m $ 34\,$s \\
     $\omega_0=\omega$ &
     $(t,t')$ &  $3.3\cdot 10^{-11}$ & $ 1.6\cdot 10^{-12}$  &   $21\,$
     m $ 14\,$s \\
     & RK4 &  $9.4\cdot 10^{-8}$ & $ 1.7\cdot 10^{-13}$ &  $38\,$
     m $ 24\,$s\\
     \hline
   \end{tabular}
   \caption{Comparison of highly accurate methods. 
   }
  \label{tab:erritottprime}
\end{table}
For a moderate time-dependence, i.e. in the 
case of the rotating-wave approximation ($\omega_0 = 0$), the $(t,t')$ 
method and the Chebychev propagator with iterative time ordering yield
a similarly good performance in terms of errors and CPU time.
Contradicting the common perception of the Runge-Kutta scheme as a
particularly efficient method, the CPU time for our example is found
to be almost two orders of magnitude and the error three orders of
magnitude larger than for the Chebychev propagator with iterative
time ordering and the $(t,t')$ method.

For strong time-dependence, i.e. resonant driving without the
rotating-wave approximation ($\omega_0 = \omega$), the Chebychev
propagator with iterative time ordering is  found by far superior in terms of
both efficiency and accuracy compared to the $(t,t')$ method and the
fourth-order Runge-Kutta scheme.
In both cases, the Runge-Kutta scheme is the least accurate method.
The smallest error of the solution, $\mathrm{\varepsilon_{sol}^{\max}}$,
achieved with RK4 is of the order of $10^{-7}$ for $\omega_0 = \omega$
and $\Delta t = 10^{-6}\,$a.u. and of the order of $10^{-9}$ for
$\omega_0 = 0$ with the same  $\Delta t$.
We have not tested smaller time steps, since already with
$\Delta t = 10^{-6}\,$a.u.,  RK4 is the least efficient of the three
methods in terms of CPU time.

\begin{figure}[btp]
  \centering
  \includegraphics[width = 0.95\linewidth]{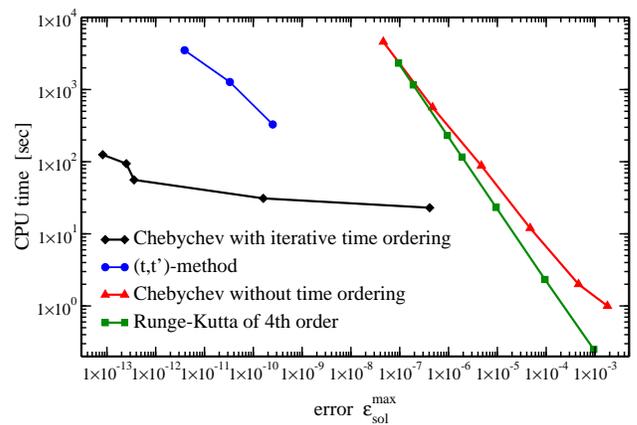}
  \caption{ (color online)
    Comparison of propagation methods for strongly time-dependent
    Hamiltonian ($\omega_0=\omega$)
    in terms of the CPU time required in order not
    to exceed a given maximum error of the solution,
    $\mathrm{\varepsilon_{sol}^{\max}}$.
  }
  \label{fig:error_vs_cpu}
\end{figure}

Figure \ref{fig:error_vs_cpu} illustrates how much 
CPU time is required for a given maximum error of the
solution, $\mathrm{\varepsilon_{sol}^{\max}}$.
A clear separation between highly accurate methods (Chebychev
propagator with iterative time ordering,  $(t,t')$ method) 
and less accurate methods (standard Chebychev propagator without
time ordering, fourth-order Runge-Kutta scheme) emerges. 
If a highly accurate method is desired, the Chebychev
propagator with iterative time ordering appears to be the best
choice. It outperforms the $(t,t')$ method not only in terms of CPU
time as shown in Fig.~\ref{fig:error_vs_cpu} but also in terms of
required memory.
 In the intermediate regime realizing a comprise
between accuracy and efficiency, the Chebychev propagator with
iterative time ordering is still the best choice. While the less accurate
methods that ignore time ordering become prohibitively expensive, the
$(t,t')$ method does not cover this regime. This is due to the choice
of $N_{t'}$ -- if it is large enough, the calculation is converged
and the error is very small, if it is too large, convergence cannot be
achieved and norm conservation is violated.
Only for cases, where a limited accuracy of the solution
is sufficient ($\varepsilon_\mathrm{sol}>10^{-5}$), the standard Chebychev
propagator and the fourth-order Runge-Kutta scheme represent the most 
efficient propagation schemes.

\subsection{Wave packet interferometry}
\label{subsec:example3} 
Our third example applies the Chebychev propagator with iterative
time ordering to a model that cannot be integrated analytically. It
explores the effect of time ordering on phase sensitivity as employed
in coherent control. 
Wave packet interferometry has first been demonstrated in the
early 1990s.\cite{schererJCP91} 
A pair of electronic or vibrational wave packets are made to interfere 
by two laser pulses. This represents a conceptually very simple
prototype of quantum control.\cite{Tannorbook}
The interference is controlled by the relative phase between the two
pulses.

\begin{figure}[tbp]
  \centering
  \includegraphics[width = 0.95\linewidth]{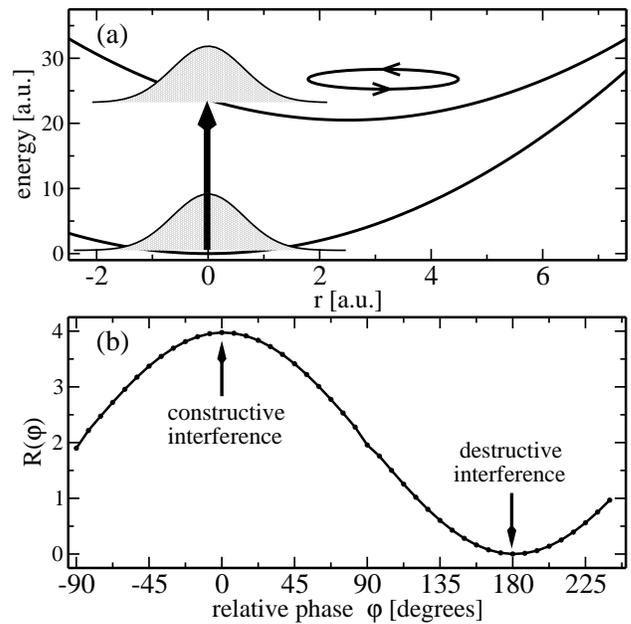}
  \caption{
   (a) Schematic representation of the generation of wave packets in the
   excited state with an ultrashort laser pulse.  
   (b) Ratio of the population on the excited state as a function of the
   relative phase between the pump and the control pulse. $t_1$ is the end of the
   pump pulse.
  }
  \label{fig:wpi}
\end{figure}

Our example is inspired by a recent experiment.\cite{OhmoriPRL06}
We consider two harmonic oscillators that are coupled by a laser
field, 
\begin{equation}
\Op{H} =
\begin{pmatrix}
  \Op{T}+ \Op{V}_g(r) & \Op{\mu} E(t) \\
    \Op{\mu} E(t) &  \Op{T}+ \Op{V}_e(r)  \\
    \end{pmatrix}  \,,
\end{equation}
where $\Op{T}$ denotes the kinetic energy and 
\begin{eqnarray*}
\Op{V}_g(r) &=& \frac{1}{2m} \omega_g^2 r^2\,, \\
\Op{V}_e(r) &=& \frac{1}{2m} \omega_e^2 (r-r_e)^2\,.
\end{eqnarray*}
For simplicity we again take $m=1$, $\omega_g = \omega_e= 1$, and $\mu=1\,$a.u.
The Hamiltonian is represented on a Fourier grid with
$N_{\mathrm{grid}} =128$, $r_{\min} = -10\,$a.u., $r_{\max} = 12\,$a.u. and 
$r_e = 3.5\,$a.u.
Starting from the vibronic ground state, 
a pump pulse is applied to create a wave packet in the excited state,
cf. Fig.~\ref{fig:wpi}a. The excited state wave packet oscillates back
and forth in the excited state potential with a period of
$2\pi/\omega_e$. 
The control pulse, with parameters identical to those of the pump
pulse, can be applied with different time delays. If it is
applied after one vibrational period,
a relative phase equal to zero  induces constructive
interference while a relative phase of $\pi$ induces destructive
interference.\cite{Tannorbook} Different time delays combined with a
different choice of the relative phase yield the same
result.\cite{Tannorbook} Constructive interference implies  an
increase of population in the excited state, while for destructive
interference  the wave packet is deexcited to the ground state.

The excited state population that was measured in the experiment by a probe
pulse,\cite{OhmoriPRL06} can be simply calculated,
$|\langle\psi_e|\psi_e\rangle|^2$.
The ratio of excited state population at the final time $T$
and at time $t_1$, just after the pump pulse, 
\begin{equation}
R(\varphi) = \frac{|\langle\psi_e(T)|\psi_e(T)\rangle|^2}
{|\langle\psi_e(t_1)|\psi_e(t_1)\rangle|^2}\,,
\label{eq:ratio}
\end{equation}
depends on the relative phase between the two pulses, $\varphi$.
This dependence is illustrated in Fig. \ref{fig:wpi}b.
For $\varphi = 0$, the population increases by a factor of four and for
$\varphi = \pi$ complete de-excitation is observed. 

\begin{figure}[tbp]
  \centering
  \includegraphics[width = 0.95\linewidth]{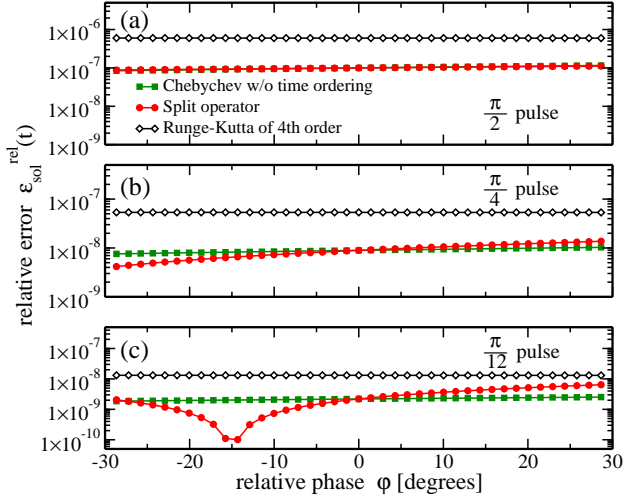}
  \caption{
    (color online) Constructive wave packet interference:
    Accuracy of the standard Chebychev propagator, the split
    propagator and the 4th order Runge-Kutta scheme    with
    respect to the Chebychev propagator with iterative time ordering for 
    different pulse areas.
  }
  \label{fig:dev_cheb_from_ito_cons}
\end{figure}

\begin{figure}[tbp]
  \centering
  \includegraphics[width = 0.95\linewidth]{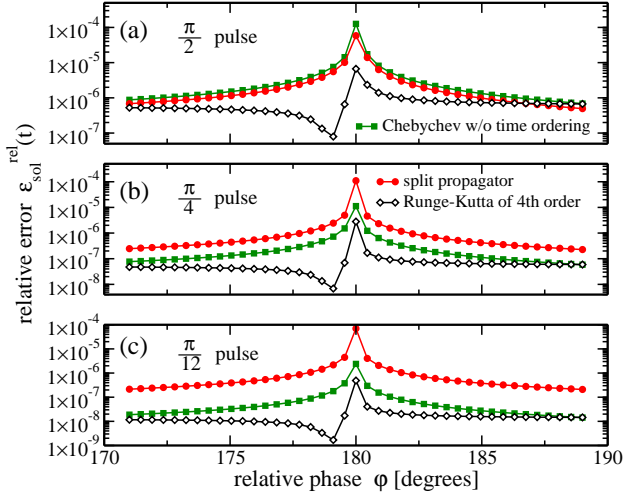}
  \caption{
    (color online) Destructive wave packet interference:
    Accuracy of the standard Chebychev propagator, the split
    propagator and the 4th order Runge-Kutta scheme    with
    respect to the Chebychev propagator with iterative time ordering for 
    different pulse areas.
  }
  \label{fig:dev_cheb_from_ito_des}
\end{figure}

Since an analytical solution is not available for this example, we
take the solution obtained by the Chebychev propagator with iterative
time ordering as the reference.
The accuracy of propagators without
time ordering is analyzed in terms of the relative
error $\varepsilon_{sol}^{rel}$,
\begin{equation}
\varepsilon_{sol}^{rel}(\varphi) =
\frac{|R_{ITO}(\varphi) - R(\varphi)|}{R_{ITO}(\varphi)}\,.
\label{eq:epsilon_relative}
\end{equation}
They are shown for the standard Chebychev propagator, the split
propagator and the fourth-order Runge-Kutta scheme 
in Figs.~\ref{fig:dev_cheb_from_ito_cons} and
\ref{fig:dev_cheb_from_ito_des}
for different pulse energies (respectively, pulse areas)
and $\Delta t=10^{-4}\,$a.u. (which has to be compared to the duration
of the pulse, $0.3\,$a.u. and the vibrational period, $2\pi\,$a.u.).
Fig.~\ref{fig:dev_cheb_from_ito_cons}
corresponds to (almost) constructive interference,
$\varphi\approx 0$,
Fig.~\ref{fig:dev_cheb_from_ito_des}
to (almost) destructive
interference, $\varphi\approx \pi$.
Overall, the relative errors obtained are smaller for wave packet
interference compared to the examples of the previous
sections~\ref{subsec:example1} and \ref{subsec:example2}.
We attribute this to the fact that the pump and control pulse are very
short compared to the vibrational time scale of the oscillators. In
this regime of impulsive excitation, 
the pulses act almost as $\delta$-functions, and there is not enough
time to accumulate large errors due to neglected time ordering.
However, even in this regime the errors are non-negligible.
As expected the errors become larger with increasing pulse intensity.
The Runge-Kutta scheme yields similar errors for both constructive and
destructive interference. The results obtained by the standard
Chebychev propagator without time-ordering and the split propagator 
appear to be only weakly affected by time ordering  for constructive
inteference. 
The results obtained with these two propagators
for destructive interference are much more sensitive to
time ordering effects and relative errors reach between 
$10^{-6}$ and $10^{-4}$  for weak and strong pulses, respectively. The
error of the split propagator is due to two effects -- time-ordering
and the non-vanishing commutator between kinetic and potential energy
while the error of the standard Chebychev propagator is solely due to
time ordering. For weak pulses, the standard Chebychev propagator
yields more accurate results than the split propagator, cf.
Fig.~\ref{fig:dev_cheb_from_ito_des}. However, for
strong pulses (pulse area of $\pi/2$) roughly the same accuracy is
achieved by the standard Chebychev propagator and the split
propagator. This indicates that the neglected time-ordering becomes
the dominating source of error.


\section{Conclusions}
\label{sec:concl}

We have developed a Chebychev propagator based on iterative
time ordering to solve TDSEs with an explicitly time-dependent
Hamiltonian. The key idea consists in rewriting the term of the TDSE that
contains the time-dependence of the Hamiltonian as an
inhomogeneity. This inhomogeneity can be approximated iteratively. At 
each step of the iteration, the Chebychev propagator for inhomogeneous
Schr\"odinger equations\cite{NdongJCP09} is employed. Convergence is
reached when the wave functions of two consecutive iteration
steps differ by less than a pre-specified error. Time ordering is thus
accomplished in an implicit manner.

We have outlined the implementation of the algorithm and demonstrated
the accuracy and efficiency of this propagator for three different
examples. A comparison to analytical solutions and other available
propagators has shown our approach to be extremely accurate, yet
efficient, in particular for very strong time dependencies.

The importance of correctly accounting
for time ordering effects\cite{KormannJCP08}
has been demonstrated for destructive
quantum interference phenomena. In most of the literature on coherent
control, accurate propagation methods are employed but time ordering
effects are completely neglected. This is not justified, in particular
for applications such as optimal control theory or high-harmonic
generation where the fields are very strong. 

The approach of rewriting parts of the TDSE as an inhomogeneity can be
extended to other classes of problems where numerical integration is
difficult. An obvious example is given by non-linear Schr\"odinger
equations such as the Gross-Pitaevski equation. By rewriting the
non-linear term as an inhomogeneity, it should be possible to derive a
very stable propagation scheme. 

\begin{acknowledgments}
  We would like to thank Jos\'e Palao and Mathias Nest for fruitful
  discussions. Financial support from the Deutsche
  Forschungsgemeinschaft through Sfb 450 (MN, RK) and an Emmy-Noether
  grant (CPK) are gratefully acknowledged.
\end{acknowledgments}


\appendix
\section{Transformation to obtain the coefficients
  $|\Phi^{(j')}\rangle$ from the Chebychev expansion coefficients
  $|\bar\Phi_j\rangle$ of the inhomogeneous term}  
\label{app:trafo}

In order to make use of Eq.~(\ref{eq:monic}),
a transformation linking the  Chebychev coefficients, 
$|\bar{\Phi}_j\rangle$ that are calculated by cosine transformation of
the inhomogeneous term, cf. Eq.~(\ref{eq:barphisum}),
to the coefficients $|\Phi^{(j')}\rangle$
appearing in the formal solution of the inhomogeneous Schr\"odinger
equation, cf. Eqs.~(\ref{eq:formalsolhil})-(\ref{eq:Fm}), is required.
Assuming $\tau \in [0,t]$, then $\bar\tau = 2\tau/t - 1$, and
Eq.~(\ref{eq:monic}) becomes
\begin{equation}
\sum_{j=0}^{m-1}P_j(\bar \tau)|\bar \Phi_j\rangle = 
\sum_{j'=0}^{m-1} \frac{\tau^{j'}}{j'!}|\Phi^{(j')}\rangle\,.
\label{eq:barphitophi}
\end{equation}
Replacing $\tau$ by $\bar \tau$ in the right-hand side of
(\ref{eq:barphitophi}), one obtains
\begin{equation}
\sum_{j=0}^{m-1}P_j(\bar \tau)|\bar \Phi_j\rangle = 
\sum_{j'=0}^{m-1} \frac{(\bar \tau+1)^{j'} t^{j'}}{j'!2^{j'}}|\Phi^{(j')}\rangle\,.
\label{eq:coefpoly}
\end{equation}
The Chebychev polynomials can be expanded in powers of $\bar\tau$,
\begin{equation}
P_j(\bar \tau) = \sum_{k=0}^{j} C_{j,k}\frac{\bar \tau^k}{k!}  \,.
\label{eq:chebpoly}
\end{equation}
Since Chebychev polynomials satisfy
\begin{equation}
P_{j+1}(\bar \tau) = 2\bar \tau P_j(\bar \tau) - P_{j-1}(\bar \tau)\,,
\end{equation}
the coefficients $C_{j,k}$ satisfy a corresponding recursion relation,
cf. Eq.~(A6) of Ref.~\onlinecite{NdongJCP09}. 
Inserting Eq.~(\ref{eq:coefpoly}) into  Eq.~(\ref{eq:barphitophi})
yields 
\begin{equation}
  \sum_{j=0}^{m-1}  \sum_{k=0}^{j} \frac{C_{j,k}}{k!} |\bar
\Phi_j\rangle \bar \tau^k  = 
\sum_{j'=0}^{m-1}  \sum_{k=0}^{j'}  \frac{j'!} {k!(j'-k)!}\frac{ 
t^{j'}}{j'!2^{j'}}|\Phi^{(j')}\rangle \bar \tau^k \,.
\label{eq:coefchebytocoefpoly}
\end{equation}
Introducing
\begin{eqnarray*}
  |\bar \alpha_{j,k}\rangle  &=& \frac{C_{j,k}}{k!} |\bar\Phi_j\rangle
  \,, \\
  |\beta_{j',k}\rangle &=&  \frac{1} {k!(j'-k)!}\frac{
  t^{j'}}{2^{j'}}|\Phi^{(j')}\,, 
\end{eqnarray*}
Eq.~(\ref{eq:coefchebytocoefpoly}) is rewritten 
\begin{equation}
 \sum_{j=0}^{m-1}  \sum_{k=0}^{j} |\bar \alpha_{j,k}\rangle \bar \tau^k =
     \sum_{j'=0}^{m-1}  \sum_{k=0}^{j'}  |\beta_{j',k}\rangle \bar \tau^k \,.
\label{eq:coefchebytocoefpoly1}
\end{equation}
Calculation of the $|\Phi^{(j')} \rangle$ from the $|\bar \Phi_j\rangle$  
is thus equivalent to calculate the $|\beta_{j',k} \rangle$ from the $|\bar \alpha_{j,k}
\rangle$. Note that the powers of $\bar \tau$ in Eq.~(\ref{eq:coefchebytocoefpoly1}) 
occur in the inner sums. In order to transform them to the outer sums, 
first the left-hand side of Eq.~(\ref{eq:coefchebytocoefpoly}) is  written 
\begin{widetext}
\begin{eqnarray}
\label{eq:chebdetail1}
   \sum_{j=0}^{m-1}  \sum_{k=0}^{j} \frac{C_{j,k}}{k!} |\bar
\Phi_j\rangle \bar \tau^k &=& 
|\bar\alpha_{0,0}\rangle + \sum_{k=0}^{1}
|\bar \alpha_{1,k}\rangle \bar \tau^k + 
\sum_{k=0}^{2} |\bar \alpha_{2,k}\rangle \bar \tau^k + \cdots +
\sum_{k=0}^{m-1} |\bar \alpha_{m-1,k}\rangle \bar \tau^k \,,  \\
\sum_{j=0}^{m-1}  \sum_{k=0}^{j}  \frac{C_{j,k}}{k!} |\bar
\Phi_j\rangle \bar \tau^k &=& 
\sum_{j=0}^{m-1}| \bar \alpha_{j,0}\rangle + 
\sum_{j=1}^{m-1} |\bar \alpha_{j,1}\rangle \bar \tau  +
\sum_{j=2}^{m-1} |\bar \alpha_{j,2}\rangle \bar \tau^2 + \cdots +
|\bar \alpha_{m-1,m-1}\rangle  \bar \tau^{m-1}\,, \\
\label{eq:chebdetail2}
   \sum_{j=0}^{m-1}  \sum_{k=0}^{j} \frac{C_{j,k}}{k!} |\bar
\Phi_j\rangle  \bar \tau^k &=&
\sum_{j=0}^{m-1}  \sum_{k=j}^{m-1} |\bar \alpha_{j,k}\rangle  \bar \tau^j\,. 
\label{eq:chebdetail3}
\end{eqnarray}
\end{widetext}
Similarly, the right-hand side of Eq.~(\ref{eq:coefchebytocoefpoly}) is
written 
\begin{equation}
   \sum_{j'=0}^{m-1}  \sum_{k=0}^{j'}  \frac{1} {k!(j'-k)!}\frac{
t^{j'}}{2^{j'}}|\Phi^{(j')}\rangle   \bar \tau^k = 
\sum_{j'=0}^{m-1}  \sum_{k=j'}^{m-1} |\beta_{j',k}\rangle  \bar \tau^{j'} \,.
\label{eq:polydetail2}
\end{equation}
Equating the right-hand sides of Eq.~(\ref{eq:chebdetail2}) and
Eq.~(\ref{eq:polydetail2}), the $|\beta_{j',k} \rangle$ are obtained,
\begin{equation}
\sum_{j'=k}^{m-1}|\beta_{j',k}\rangle  =  \sum_{j=k}^{m-1}
|\bar \alpha_{j,k}\rangle, \,\,\,\, 0\le k \le m-1 \,. 
\label{eq:beta}
\end{equation}
\begin{widetext}
Replacing $|\bar \alpha_{j,k}\rangle$ and $|\beta_{j',k}\rangle$ by their
definition yields 
\begin{eqnarray}
  \sum_{j'=k}^{m-1} \frac{1} {k!(j'-k)!}\frac{ t^{j'}}{2^{j'}}|\Phi^{(j')}\rangle  &=&
\sum_{j=k}^{m-1}
\frac{C_{j,k}}{k!} |\bar\Phi_j\rangle , \qquad 0 \le k \le m-1 \,, \nonumber \\
\frac{1}{k!} \sum_{j'=k}^{m-1} \frac{1} {(j'-k)!}\frac{
  t^{j'}}{2^{j'}}|\Phi^{(j')}\rangle  &=& 
 \frac{1}{k!}  \sum_{j=k}^{m-1} C_{j,k} |\bar\Phi_j\rangle , \qquad  0 \le k \le m-1 \,,\\
\sum_{j'=k}^{m-1} \frac{1} {(j'-k)!}\frac{ t^{j'}}{2^{j'}}|\Phi^{(j')}\rangle  &=&
 \sum_{j=k}^{m-1} C_{j,k} |\bar\Phi_j\rangle \,, \qquad  0 \le k \le m-1\,.
\nonumber
\end{eqnarray}
This leads to the hierarchy of equations
\begin{eqnarray}
  \frac{ t^{m-1}}{2^{m-1}}|\Phi^{(m-1)}\rangle &=&  
C_{m-1,m-1}
|\bar\Phi_{m-1}\rangle   \,, \nonumber \\
 \frac{ t^{m-2}}{2^{m-2}}|\Phi^{(m-2)}\rangle
+ \frac{ t^{m-1}}{2^{m-1}}|\Phi^{(m-1)}\rangle &=& 
C_{m-2,m-2} |\bar\Phi_{m-2}\rangle + 
C_{m-1,m-2} |\bar\Phi_{m-1}\rangle \,, \nonumber \\ 
\vdots &=& \vdots         \nonumber \\
\sum_{j'=k}^{m-1} \frac{1} {(j'-k)!}\frac{
  t^{j'}}{2^{j'}}|\Phi^{(j')}\rangle  &=&  \sum_{j=k}^{m-1} 
C_{j,k} |\bar\Phi_j\rangle , \qquad 0 \le k \le m-2 \,.
\end{eqnarray}  
The coefficients $|\Phi^{(j')} \rangle$ can thus be determined step by step
from the coefficients of the Chebychev expansion, $|\bar\Phi_{j}\rangle$, 
\begin{eqnarray}
|\Phi^{(m-1)}\rangle &=&  
\frac{ 2^{m-1}}{t^{m-1}} C_{m-1,m-1}
|\bar\Phi_{m-1}\rangle \,,
\label{eq:barphitophiend1}\\
 |\Phi^{(k)}\rangle &=&  \frac{ 2^{k}}{t^{k}} \left( 
\sum_{j=k}^{m-1} C_{j,k} |\bar\Phi_j\rangle  -  \sum_{j=k+1}^{m-1} 
 \frac{1} {(j'-k)!}\frac{ t^{j'}}{2^{j'}}|\Phi^{(j')}\rangle \right )\,, \qquad k = m-2,
0\,.
\label{eq:barphitophiend2}
\end{eqnarray}  
\end{widetext}
Note that the transformation given by Eqs.~(\ref{eq:barphitophiend1})
and (\ref{eq:barphitophiend2}) becomes numerically instable for large
orders, $m \sim 100$. In our applications, such a large $m$ would
correspond to time steps larger than the overall propagation time and 
was never required. 

%

\end{document}